\documentclass[prd,eqsecnum,preprintnumbers,twocolumn,showpacs]{revtex4}
\usepackage{amsmath,bm,graphicx,slashed}

\newcommand{\3}[1]{\bm{#1}}

\begin{document}

\preprint{ANL-HEP-PR-07-25}
\title{$k_T$-factorization is violated in production of
  high-transverse-momentum particles in hadron-hadron collisions} 

\author{John Collins}
\email{collins@phys.psu.edu}
\affiliation{
    Physics Department,
    Penn State University,
    104 Davey Laboratory,
    University Park PA 16802,
    U.S.A.
}

\author{Jian-Wei Qiu}
\email{jwq@iastate.edu}
\affiliation{
    Department of Physics and Astronomy,
    Iowa State University,
    Ames IA 50011,
    U.S.A.
}
\affiliation{
    High Energy Physics Division, 
    Argonne National Laboratory,
    Argonne IL 60439,
    U.S.A.
}

\date{28 June 2007}

\begin{abstract}
  We show that hard-scattering factorization is violated in the
  production of high-$p_T$ hadrons in hadron-hadron collisions, in the
  case that the hadrons are back-to-back, so that $k_T$ factorization
  is to be used.  The explicit counterexample that we construct is for
  the single-spin asymmetry with one beam transversely polarized.  The
  Sivers function needed here has particular sensitivity to the Wilson
  lines in the parton densities.  We use a greatly simplified model
  theory to make the breakdown of factorization easy to check
  explicitly.  But the counterexample implies that standard arguments
  for factorization fail not just for the single-spin asymmetry but
  for the unpolarized cross section for back-to-back hadron production
  in QCD in hadron-hadron collisions.  This is unlike corresponding
  cases in $e^+e^-$ annihilation, Drell-Yan, and deeply inelastic
  scattering.  Moreover, the result endangers factorization for more
  general hadroproduction processes.
\end{abstract}
\pacs{
   12.38.Bx, 
   12.39.St, 
   13.85.Ni, 
   13.87.-a, 
   13.88.+e  
}

\maketitle

\section{Introduction}

The great importance of hard-scattering factorization in high-energy
phenomenology hardly needs emphasis.  Essential to its application and
predictiveness is the universality of parton densities (and
fragmentation functions, etc) between different reactions.  However,
as can be seen from
\cite{bomhof_04,Bacchetta:2005rm,Bomhof:2006dp,Pijlman:2006tq},
process-dependent Wilson lines appear to be needed in the inclusive
production of two high-transverse-momentum particles in hadron-hadron
collisions, i.e., in the process
\begin{equation}
\label{eq:HHHH}
  H_1 + H_2 \to H_3 +H_4 + X.
\end{equation}
In this paper we will show that this situation definitively leads to a
breakdown of factorization.  

The standard expectation is that the cross section is a convolution of
a hard scattering coefficient $d\hat\sigma$, parton densities,
fragmentation functions and a possible soft function:
\begin{align}
\label{eq:fact}
  E_3E_4 \frac{d\sigma}{ d^3\3{p}_3 d^3\3{p}_4 }
= {}&
  \sum\int d\hat\sigma_{i+j\to k+l+X} \,
   f_{i/1} \, f_{j/2} \,
   d_{3/k} \, d_{4/l}
\nonumber\\
&
   + \mbox{power-suppressed correction}.
\end{align}
Here the sum and integral are over the flavors and momenta of the
partons of the hard scattering, $f_{i/H}$ denotes a parton density,
and $d_{H/i}$ a fragmentation function.

It is noteworthy that the classic published proofs for factorization
in hadron-hadron scattering \cite{bodwin_85,collins_85_88} only
concerned the Drell-Yan process.  There are a number of difficult
issues in the proof that are highly non-trivial to extend to other
reactions in hadron-hadron collisions, even though Eq.\
(\ref{eq:fact}) is a standard expectation.

We will examine the case that the produced hadrons are almost
back-to-back.  Then the appropriate factorization property is
$k_T$-factorization, which entails \cite{TMD} the use of
transverse-momentum dependent (TMD) parton densities and fragmentation
functions.  However, the issues raised by our counterexample to
factorization are sufficiently general that they create a need to
examine very carefully the arguments for factorization in
hadroproduction of hadrons even in situations where ordinary collinear
factorization with integrated densities is appropriate.  In the case
of $k_T$-factorization with TMD densities, the factorization formula
needs the insertion of a soft factor $S$, not shown in Eq.\
(\ref{eq:fact}).

The problems concern gluon exchanges between different kinds of
collinear line, as in Fig.\ \ref{fig:HHHH1} below.  To obtain
factorization, the gluon attachments must be converted to Wilson lines
in gauge-invariant definitions of the parton densities and
fragmentation functions.  This relies \cite{collins_85_88} on the use
of Ward identities applied to approximations to the amplitudes.  But
the approximations are only valid after certain contour deformations
on the loop momenta.

Bacchetta, Bomhof, Mulders and Pijlman
\cite{bomhof_04,Bacchetta:2005rm,Bomhof:2006dp,Pijlman:2006tq} argued
that because of the complicated combination of initial- and
final-state interactions, the Wilson lines must be modified.  What is
not so clear is the interpretation of their result.  So in the present
paper we present an argument to make fully explicit the failure of
factorization.

Since the issue is one of factorization in general, and not just
specifically in QCD, we clarify the issue by examining a particular
process in a model field theory.  The process is a transverse
single-spin asymmetry of the kind controlled by a Sivers function.
This is a case where problems in the contour deformation directly
affect the value of the cross section at the lowest possible order of
perturbation theory.  Our model field theory is simple enough that the
calculations and their interpretation as implying factorization
violation are unambiguous.  But, as we will explain in the final
section, we expect the failure of factorization to be more general:
our particular process and model simply make it very easy to see the
failure.

\section{Construction of model}

Since proofs of factorization apply to quantum field theories in
general (if they are renormalizable), the construction of a
counterexample, to demonstrate and to \emph{understand} a failure of
the normal methods of proof, is conveniently done in a simple model
theory. 

Our model resembles the one used by Brodsky, Hwang, and Schmidt
\cite{BHS} in their discussion of single spin asymmetries.  It is
defined as follows:
\begin{itemize}
\item The gauge group is abelian.  This simplifies the graphs, and
  allows the next feature.
\item The gluon is massive.  This avoids the discussion being confused
  by actual infra-red divergences in the S-matrix.
\item The initial-state particles correspond to Dirac fields that are
  neutral under the gauge group.  We will call them hadrons.  The
  fields need to be Dirac fields in order to have the single
  transverse spin asymmetries that we will examine.  We will use two
  types of hadron.
\item Each ``hadron field'' $H_i$ will have a coupling to a Dirac
  field $\psi_i$ and a scalar field $\phi_i$, which in \cite{BHS} would be
  called a diquark field:
  \begin{equation}
    \lambda_i \left( \bar{H}_i\psi_i \, \phi_i^\dag{} + \bar{\psi}_iH_i \, \phi_i \right). 
  \end{equation}
  The quark field $\psi_i$ has a coupling $g_i$ to the gauge field, and
  the scalar field $\phi_i$ has the opposite coupling.  
\item All the masses in the theory are comparable, to avoid confusing
  the calculation with logarithmic dependence on large ratios of
  masses. 
\end{itemize}

In our analog of hadroproduction, Fig.\ \ref{fig:HHHH1}, the two
initial state hadrons are of the two different types, and we use
this to simplify our argument for non-factorization.  The lines in the
the lower part of graphs are chosen to be those of gauge coupling
$g_1$, while the lines in the upper part of the graph are those with
coupling $g_2$.  But if the attachments of the gluon to the upper
lines were in some way to correspond to a Wilson line in the parton
density in the hadron in the lower part of the graph, the charge would
have to be $g_1$.  Since $g_1$ and $g_2$ are arbitrary, there is no
way to make a correspondence between the graph and the Wilson line
formalism, provided that the contribution is non-zero, as we will
demonstrate.  This will also insulate us against sign errors and the
like.

We will also choose the detected outgoing particles $H_3$ and $H_4$ to
correspond to the scalar fields.  The sole purpose here is to simplify
the Dirac algebra slightly, thereby making the calculations more
transparent and elementary.

One feature of our counterexample appears to be very special to an
abelian gauge theory.  This is that the two couplings $g_1$ and $g_2$
need have no relation to each other: there is a continuous infinity of
representations of the gauge group.  In contrast, there is a single
value of the coupling $g$ for all the fields in a non-abelian theory.
The role of the ratio $g_2/g_1$ is now taken over by the
representation matrices for the different fields in QCD (triplet,
antitriplet, octet), with the different couplings related by factors
of rational numbers.  So in any particular example there is a
potential for a numerical coincidence between the sizes of the
numerical values of the graphs, which could then appear to give
consistency with factorization.  A counterexample to factorization
would then be more complicated, with a comparison of cases with
different kinds of partons (quarks, antiquarks or gluons), cf.\
\cite{bomhof_04,Bacchetta:2005rm,Bomhof:2006dp,Pijlman:2006tq}.

\section{Review of SIDIS and DY}

We now review how \cite{BHS} a transverse single-spin asymmetry (SSA)
arises in semi-inclusive deep-inelastic scattering (SIDIS), at the
level of one-gluon-exchange, and how it determines \cite{collins_02}
the Wilson line that defines parton densities.  Then we review the
differences that give factorization with an exact sign reversal in the
Sivers function for the Drell-Yan process \cite{collins_02,BHS.DY}.
This will give us methods of calculation that will give us a very
elementary way to obtain the SSA for the process (\ref{eq:HHHH}).

\subsection{SIDIS}

With the electromagnetic part of the scattering factored out, SIDIS is
the process
\begin{equation}
  \label{eq:SIDIS}
  \gamma^*(q)+H(p) \to H'(r)+X.
\end{equation}
We use light-front coordinates in which the incoming momenta are
\begin{equation}
  p = \left( p^+, \frac{m_H^2}{2p^+}, 0_T \right),
\qquad
  q = \left( -xp^+, \frac{Q^2}{2xp^+}, 0_T \right).
\end{equation}
The detected outgoing particle is defined by a longitudinal momentum
fraction $z$ and a transverse momentum $r_T$:
\begin{equation}
  r = \left( xp^+\frac{r_T^2+m_\phi^2}{zQ^2}, \frac{zQ^2}{2xp^+}, r_T \right).
\end{equation}
We will assume that $Q$ is large and that the detected transverse
momentum $r_T$ is of order a hadronic mass scale $m$.

\begin{figure}
  \centering
    \includegraphics[scale=0.5]{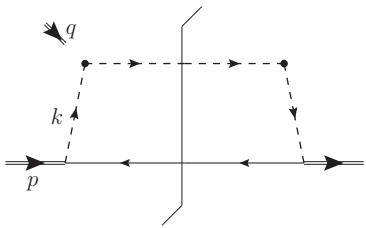}
    \caption{Lowest-order graph for SIDIS in our model.  The initial
      state particle is a color-singlet Dirac particle.  The spectator
      line is for a Dirac ``quark'' field, and the active parton is
      for a scalar ``diquark'' field, as can be enforced by a
      condition on the detected outgoing particle of momentum $q+k$.
      The arrows on the internal lines indicate the flow of color
      charge, and the vertical line cutting the graph denotes the
      final state.  }
  \label{fig:SIDIS}
\end{figure}

The lowest-order graph Fig.\ \ref{fig:SIDIS} gives the following
contribution to the differential structure tensor:
\begin{widetext}
\begin{align}
  \frac{dW^{\mu\nu}}{dz\, d^2r_T}
={}&
  \frac{\lambda^2}{4\pi}
  \int \frac{ dk^+ \, dk^- }{ (2\pi)^4 }  
    \delta\!\left( z - \frac{k^-+q^-}{q^-}\right)
    \frac{ (2k^\mu+q^\mu)(2k^\nu+q^\nu) }{ (k^2-m_\phi^2)^2 }
    (2\pi)^2 \delta\bigl((q+k)^2-m_\phi^2\bigr) \, \delta\bigl((p-k)^2-m_\psi^2\bigr) 
\times \nonumber\\& \times 
    \frac{1}{2} \textrm{Tr}\!\left[ 
       ( \slashed{p} + m_H ) \, (1+\gamma_5\slashed{s}) \,
       ( \slashed{p} - \slashed{k} + m_\psi )
     \right] .
\end{align}
The internal partons are all collinear to the target, i.e., $k^+ \sim
p^+$, $k^- \sim m^2/p^+$, $k_T \sim m$, and to leading power in $Q$, parton
model kinematics apply, so that $k^+ \simeq xp^+$ and $z\simeq 1$.  We will
assume throughout that the spin vector $s$ corresponds to a transverse
spin (in the $(q,p)$ frame), and that it is normalized so that its
extremal value obeys $s^2=-1$.  Since there is only one initial-state
hadron, we do not bother with labels to indicate the kind of hadron
(e.g., $\lambda_1$ or $\lambda_2$).

The above formula is simply related to a parton density:
\begin{equation}
  \frac{dW^{\mu\nu}}{dz\, d^2r_T}
  = \frac{ (p^\mu -q^\mu p\cdot q / q^2 ) (p^\nu -q^\nu p\cdot q / q^2 ) }{ p \cdot q }  
    \delta(z-1) P_{\phi/H}(x,r_T)
   + \mbox{power-suppressed correction},
\end{equation}
where the parton density at lowest order is
\begin{align}
\label{eq:P0}
  P_0(x,k_T)
={}&
  \frac{\lambda^2\,x(1-x)}{16\pi^3}
    \frac{ \frac{1}{2} \textrm{Tr}\!\left[ 
              ( \slashed{p} + m_H ) \, (1+\gamma_5\slashed{s}) \,
              ( \slashed{p} - \slashed{k} + m_\psi )
            \right] }
         { \bigl[ k_T^2 + m_\phi^2(1-x) + m_\psi^2x -m_H^2x(1-x) \bigr]^2 },
\end{align}
\end{widetext}
as follows from the conventional operator definition.

There is in fact no polarization dependence in this order, i.e., the
Sivers function vanishes.  At a simple calculational level, this
occurs for two reasons.  One is that $\gamma_5$ gives a non-zero
contribution only when multiplied by at least four regular Dirac
matrices,
\begin{equation}
\label{eq:gam5.trace}
  \textrm{tr} \gamma_5 \slashed{a} \slashed{b} \slashed{c} \slashed{d}
  = 4 i \epsilon_{\kappa\lambda\mu\nu} a^\kappa b^\lambda c^\mu d^\nu,
\end{equation}
while in (\ref{eq:P0}) there are at most three.
This reason for a vanishing SSA would no longer apply in a more
complicated model or with higher order graphs.  The second reason for
the vanishing is that the trace
(\ref{eq:gam5.trace}) is imaginary, while the rest of the graph is
real, so that the contribution to a cross section must be zero.

\begin{figure*}
  \centering
  \begin{tabular}{c@{\hspace*{5mm}}c}
    \includegraphics[scale=0.5]{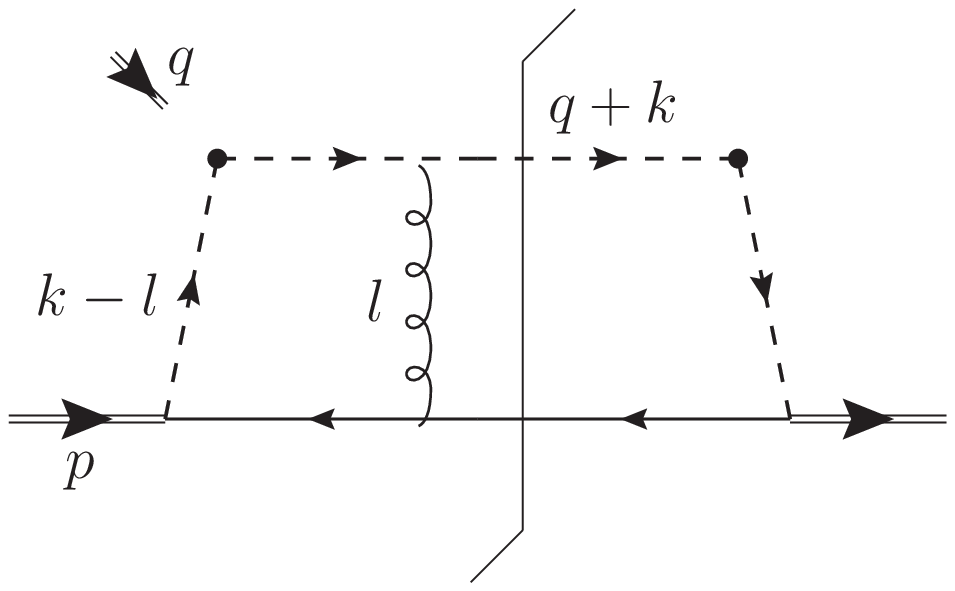}
    &
    \includegraphics[scale=0.5]{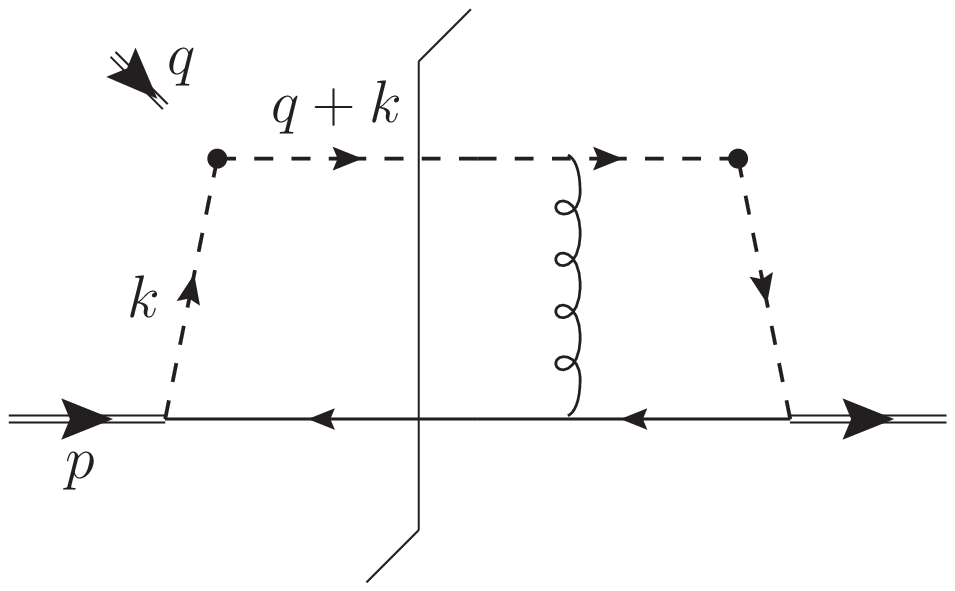}
  \\
    (a) & (b)
  \end{tabular}
  \caption{Virtual one-gluon-exchange corrections to Fig.\
    \ref{fig:SIDIS} that give a SSA.}
  \label{fig:SIDIS1}
\end{figure*}

\begin{figure}
  \centering
  \includegraphics[scale=0.5]{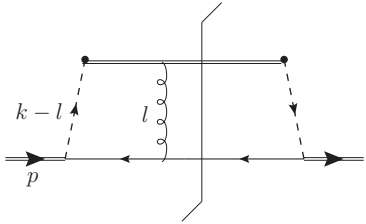}
  \caption{Virtual one-gluon-exchange correction to parton
    density. The upper double line is the Wilson line, and the graph
    shown, together with its hermitian conjugate gives the first
    contribution to the Sivers function.}
  \label{fig:pdf1}
\end{figure}

The lowest-order graphs for a non-zero SSA are the one-gluon-exchange
graphs in Fig.\ \ref{fig:SIDIS1}, which get an imaginary part from an
intermediate state that can go on-shell; this state is made by the
lines with momenta $q+k-l$ and $p-k+l$.  Standard power-counting shows
that the exchanged gluon can only be collinear to the target or soft.
The minus momentum of the gluon is trapped in the region $l^- \sim
m^2/p^+$ by the other target-collinear lines.  The on-shell
intermediate state corresponds to small angle elastic scattering, and
so to very small $l^+$, of order $p^+m^2/Q^2$.  There the only
significant dependence on $l^+$ is in the upper parton propagator.
Multiplied by the neighboring gluon vertex this gives
\begin{equation}
\label{eq:qkl}
  \frac{ -g(2q^\mu+2k^\mu-l^\mu) }{ (q+k-l)^2 -m_\phi^2+i\epsilon }
  \simeq 
  \frac{-g \delta_-^\mu}{ -l^+ + \mbox{other terms}+i\epsilon }  ,
\end{equation}
where the ``other terms'' are small or independent of $l^+$, and we
have taken a leading power approximation for the momenta in the
numerator.  The contour of integration of $l^+$ can therefore be
deformed into the lower half plane until $l$ is target-collinear.  In
that case the ``other terms'' in (\ref{eq:qkl}) are negligible, and
the denominator can be replaced by its eikonal approximation
$1/(-l^++i\epsilon)$.  This, together with a leading-power approximation in
the numerator, shows \cite{collins_02} that the gluon exchange
correction is equivalent to a contribution to the parton density with
a suitable Wilson line, Fig.\ \ref{fig:pdf1}.

Let us perform the $k^+$ and $k^-$ integrals by the on-shell
conditions on the final state, and let us perform the $l^-$
integration by contour integration.  Then the necessary imaginary part
comes simply from the imaginary part of (\ref{eq:qkl}) and thus from
the replacement
\begin{equation}
  \frac{ -g(2q^\mu+2k^\mu-l^\mu) }{ (q+k-l)^2 -m_\phi^2+i\epsilon } 
  \mapsto ig\pi\delta_-^\mu \delta(l^+).
\end{equation}
This gives rise to an SSA with the aid of the trace
\begin{multline}
    \frac{1}{2} \textrm{Tr}\!\left[ 
       ( \slashed{p} + m_H ) \, \gamma_5\slashed{s} \,
       ( \slashed{p} - \slashed{k} + \slashed{l} + m_\psi ) \,
       \gamma^+ \,
       ( \slashed{p} - \slashed{k} + m_\psi )
     \right] 
\\
\simeq 2i\epsilon_{jk} s^jl^k p^+ [m_H(1-x)+m_\psi] ,
\end{multline}
where the approximation, good to leading power, arises from the
neglect of the small components of $l$ with respect to the transverse
components. The two-dimensional $\epsilon$ tensor obeys $\epsilon_{12}=1$.  Since
the denominator in the integrand is not azimuthally symmetric in
$l_T$, the integral over $l$ gives a non-zero result for the SSA from
the whole graph.

The two graphs in Fig.\ \ref{fig:SIDIS1} are related by hermitian
conjugation and so they give equal contributions to the SSA.

\subsection{Drell-Yan}

\begin{figure}
  \centering
    \includegraphics[scale=0.45]{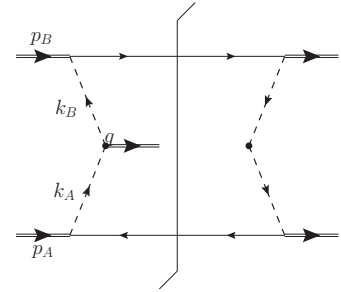}
  \caption{Lowest-order graph for the Drell-Yan process.}
  \label{fig:DY}
\end{figure}

The Drell-Yan (DY) process, 
\begin{equation}
  \label{eq:DY}
  H_A(s) + H_B \to \gamma^*(q) + X,
\end{equation}
is treated quite similarly.  We examine the cross section differential
in $q_T$, and investigate a possible SSA, with $H_A$ having a
transverse spin vector $s$.

In our model the lowest-order graph is Fig.\ \ref{fig:DY}.  It is
readily shown to be the (convolution) product of two
transverse-momentum-dependent parton densities, and, just like the
SIDIS process, it has no SSA at this order.  Note that to have the
process occur at the order shown within our model, the initial-state
hadrons $H_A$ and $H_B$ must be antiparticles of each other.

\begin{figure*}
  \centering
  \begin{tabular}{c@{\hspace*{5mm}}c}
    \includegraphics[scale=0.45]{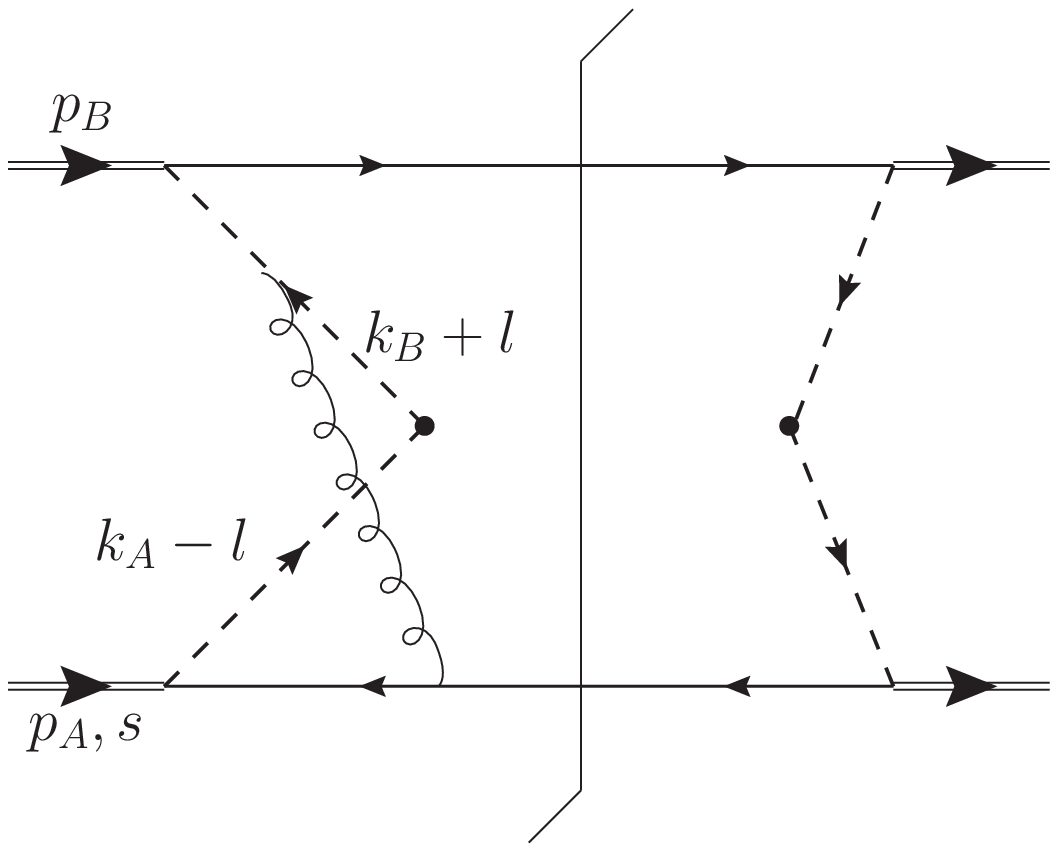}
  &
    \includegraphics[scale=0.45]{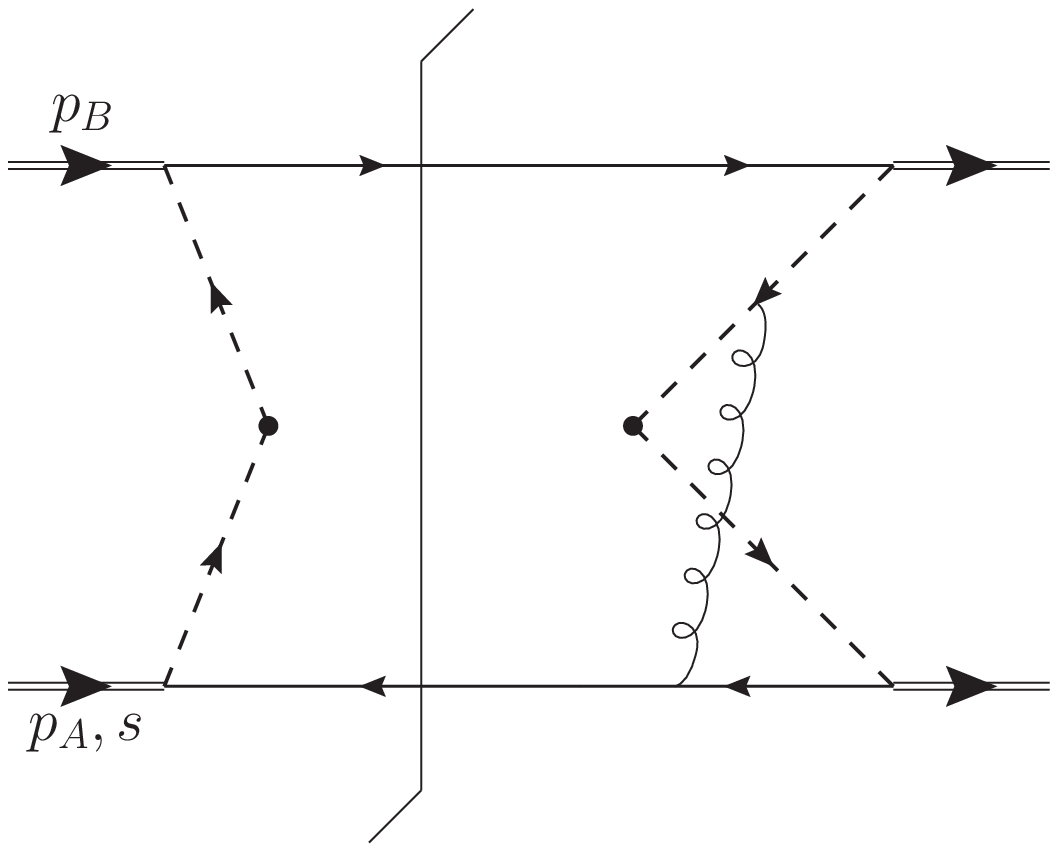}
  \\
    (a) & (b)
  \end{tabular}
  \\
  \begin{tabular}{c@{\hspace*{5mm}}c@{\hspace*{5mm}}c}
    \includegraphics[scale=0.45]{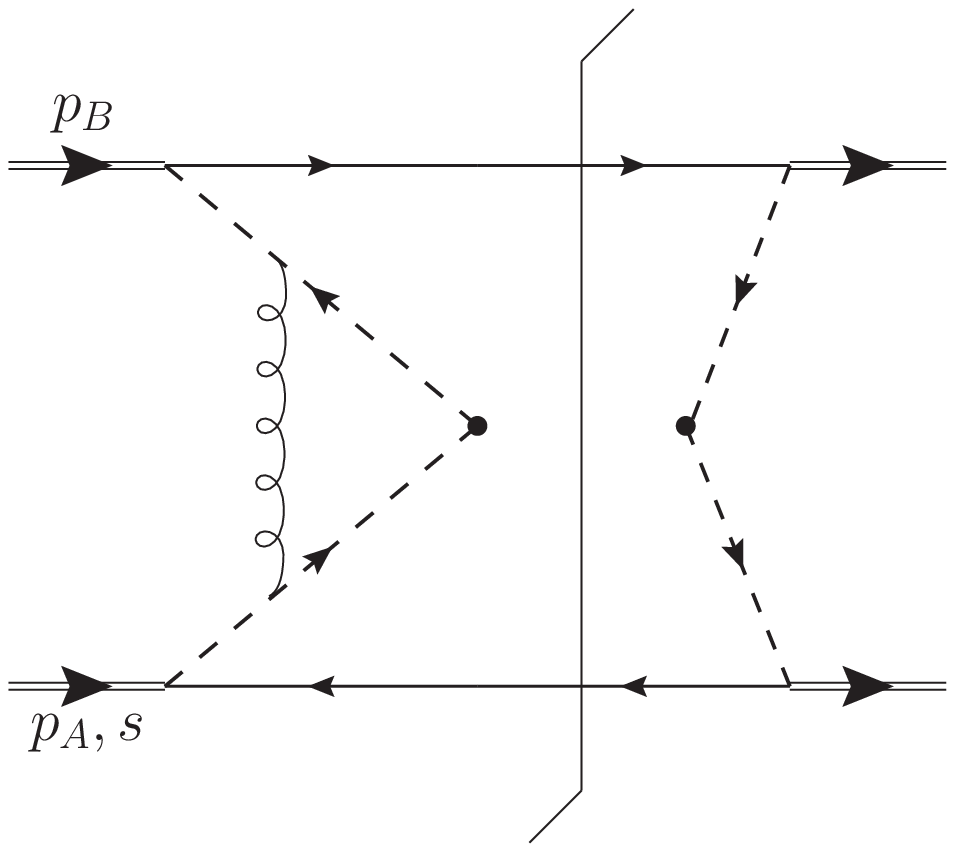}
  &
    \includegraphics[scale=0.45]{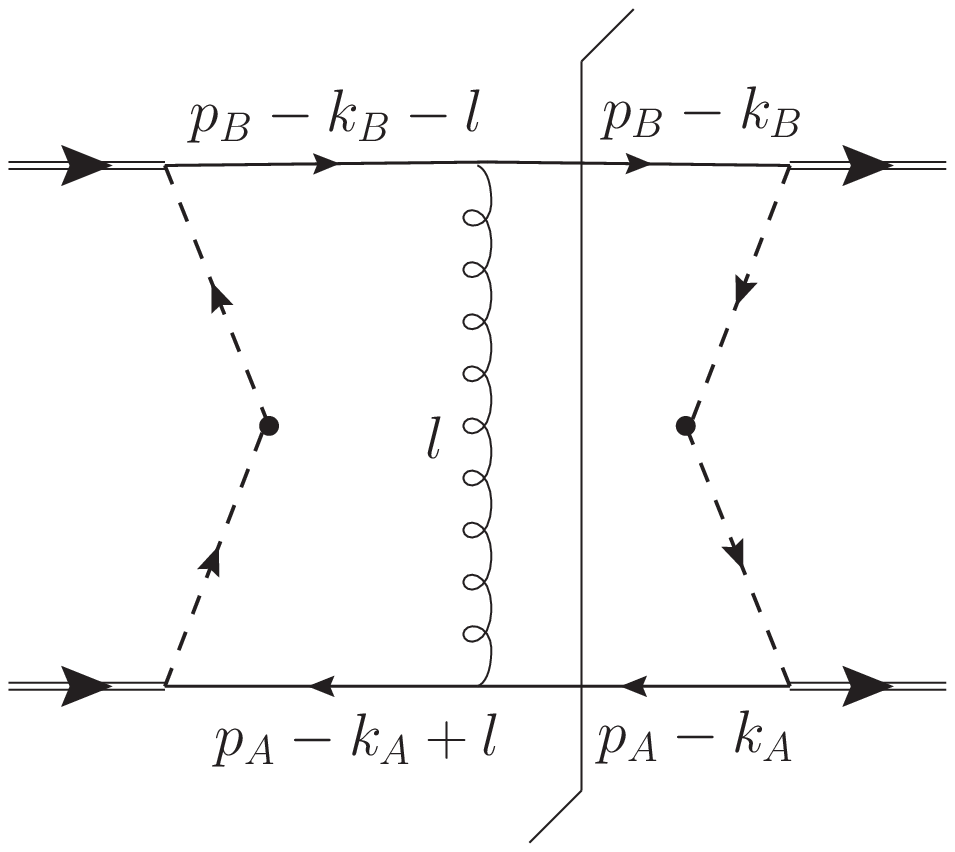}
  &
    \includegraphics[scale=0.45]{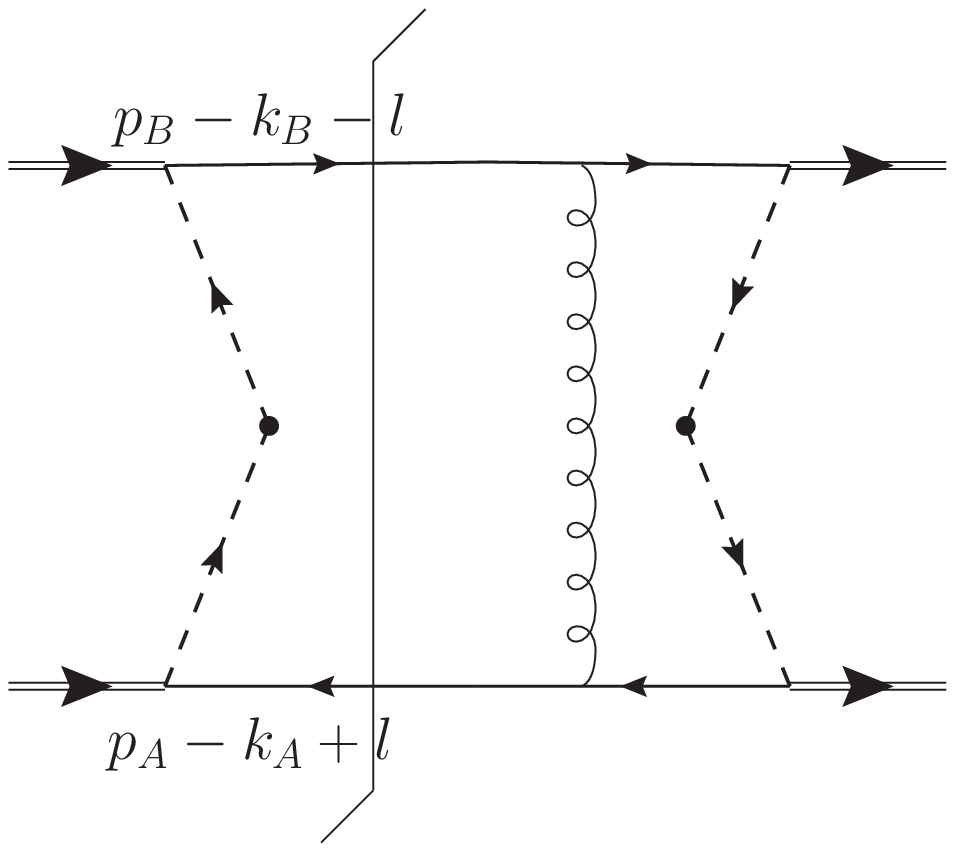}
  \\
    (c) & (d) & (e)
  \end{tabular}
  \caption{Virtual one-gluon-exchange corrections to Fig.\
    \ref{fig:DY} relevant for a SSA when the lower hadron has
    transverse spin $s$.  Graph (a) and its hermitian conjugate (b)
    have imaginary parts (at the amplitude level) that give a non-zero
    SSA.  Gluon exchange between the active partons, graph (c) and its
    not-shown conjugate, gives an imaginary part in the vertex
    correction that does not give a SSA. Spectator-spectator gluon
    exchange graphs, (d) and (e), do contribute individually to the
    SSA, but the two contributions cancel (at leading power).}
  \label{fig:DY1}
\end{figure*}

The imaginary part in the amplitude needed to get an SSA with respect
to the transverse spin $s$ of the lower hadron $H_A$ arises from
graphs such as those in Fig.\ \ref{fig:DY1}.  Graphs (a) and (b) work
just like those in Fig.\ \ref{fig:SIDIS1} for SIDIS, except that the
gluon couples to the incoming scalar antiparton instead an outgoing
scalar parton, with the necessary reversal in sign of the coupling.
Thus Eq.\ (\ref{eq:qkl}) is replaced by
\begin{equation}
  \frac{ g(2k_B^\mu+l^\mu) }{ (k_B+l)^2 -m_\phi^2+i\epsilon }
  \simeq 
  \frac{g \delta_-^\mu}{ l^+ - \mbox{other terms}+i\epsilon }  ,
\end{equation}
which gives an imaginary part exactly opposite to that for SIDIS.  The
relative sign of the $l^+$ term and the $i\epsilon$ is now that for an initial-state 
interaction, so that the Wilson line in the operator definition of the
parton density is now past-pointing instead of future-pointing
\cite{collins_02}.  As shown in \cite{collins_02}, an exact reversal
of sign of the Sivers function between SIDIS and DY follows from
the time-reversal symmetry of QCD.

No contribution to the SSA is given by graphs, like Fig.\
\ref{fig:DY1}(c), in which the both ends gluons couple the the active
partons, or where the gluon couples the \emph{upper} spectator quark
to the lower active parton.  In our model this is trivial: there are
too few Dirac matrices on the lower line to give spin-dependence.  In
a more general case, the annihilating partons could be Dirac fields,
and then the lack of spin-dependence arises because of the
eikonalization of those parton lines on the lower side of the graph
that would contribute to the imaginary part.

But spectator interaction graphs, (d) and (e), do individually
contribute to the SSA.  The imaginary part arises from putting all
four spectator lines on-shell.  Provided the gluon momentum is routed
the same way in both graphs before integration, e.g., to the left as
shown, the two contributions are equal, but exactly opposite in sign.
The cancellation is exactly one of those needed to prove factorization
\cite{collins_85_88}, and involves a sum over all allowed cuts of a
particular graph.

However, the two graphs have final states with particles of different
momenta.  This accounts for the difference compared with the
non-cancellation of the graphs in 
Fig.\ \ref{fig:SIDIS1} for SIDIS, where there is also a sum over cuts.
For the SIDIS graphs there is a requirement on the transverse momentum
of the struck parton.  In contrast, the cancellation between the two
spectator-spectator graphs in the DY process occurs because our cross
section is the fully inclusive DY cross section; no requirement was
placed on the final-state in the target fragmentation regions.  (As is
well-known from the case of diffractive hard scattering, both
theoretically \cite{Landshoff:1971zu,Henyey:1974zs} and experimentally
\cite{Abe:1997jp,Aktas:2007hn}, factorization fails when
target-relative restrictions are imposed.)

\section{Hadroproduction of hadrons}

We now have the tools to make an extremely streamlined construction of
a counterexample to factorization for the process of hadro-production
of high transverse momentum hadrons, $H_1+H_2 \to H_3+H_4+X$.  We again
use an SSA because non-factorization occurs with one gluon exchanged
beyond the lowest order in which the reaction occurs at all.  We
choose $H_1$ to be the polarized hadron, and we choose the hadrons
$H_1$ and $H_2$ to be of the two different flavors in our model.  We
also choose the detected final-state particles $H_3$ and $H_4$ to
correspond to the two flavors of scalar parton.  The
high-transverse-momentum particles $H_3$ and $H_4$ are chosen to be
almost back-to-back azimuthally (relative to the collision axis), so
that transverse-momentum-dependent parton densities and fragmentation
functions are needed for describing a factorized cross section.

\begin{figure}
  \centering
    \includegraphics[scale=0.45]{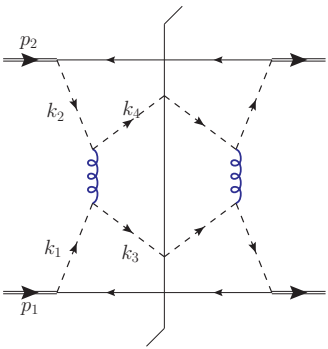}
    \caption{Lowest order graph in model for hadroproduction of
      hadrons of high transverse momentum.  The initial state
      particles are color-singlet Dirac particles.  The spectator
      lines are for Dirac ``quark'' fields of charges $g_1$ and $g_2$,
      and the active partons are for scalar ``diquark'' fields.  The
      exchanged gluon line is thickened to denote the hard scattering.
    }
  \label{fig:HHHH0}
\end{figure}

The single lowest-order graph for the process is shown in Fig.\
\ref{fig:HHHH0}.  Its hard scattering is just the gluon-exchange
subgraph.  The cross section is the convolution of the hard scattering
with a transverse-momentum-dependent parton density in each hadron.
The fragmentation functions in this order are trivial delta functions.
Although the longitudinal momenta of the incoming partons for the hard
scattering are determined from the kinematics of $H_3$ and $H_4$, only
a sum of their transverse momenta is determined.  Hence a convolution
over the transverse-momentum densities is needed.  As before, there is
no SSA at this order.

\begin{figure*}
  \centering
  \begin{tabular}{c@{\hspace*{5mm}}c@{\hspace*{5mm}}c}
    \includegraphics[scale=0.45]{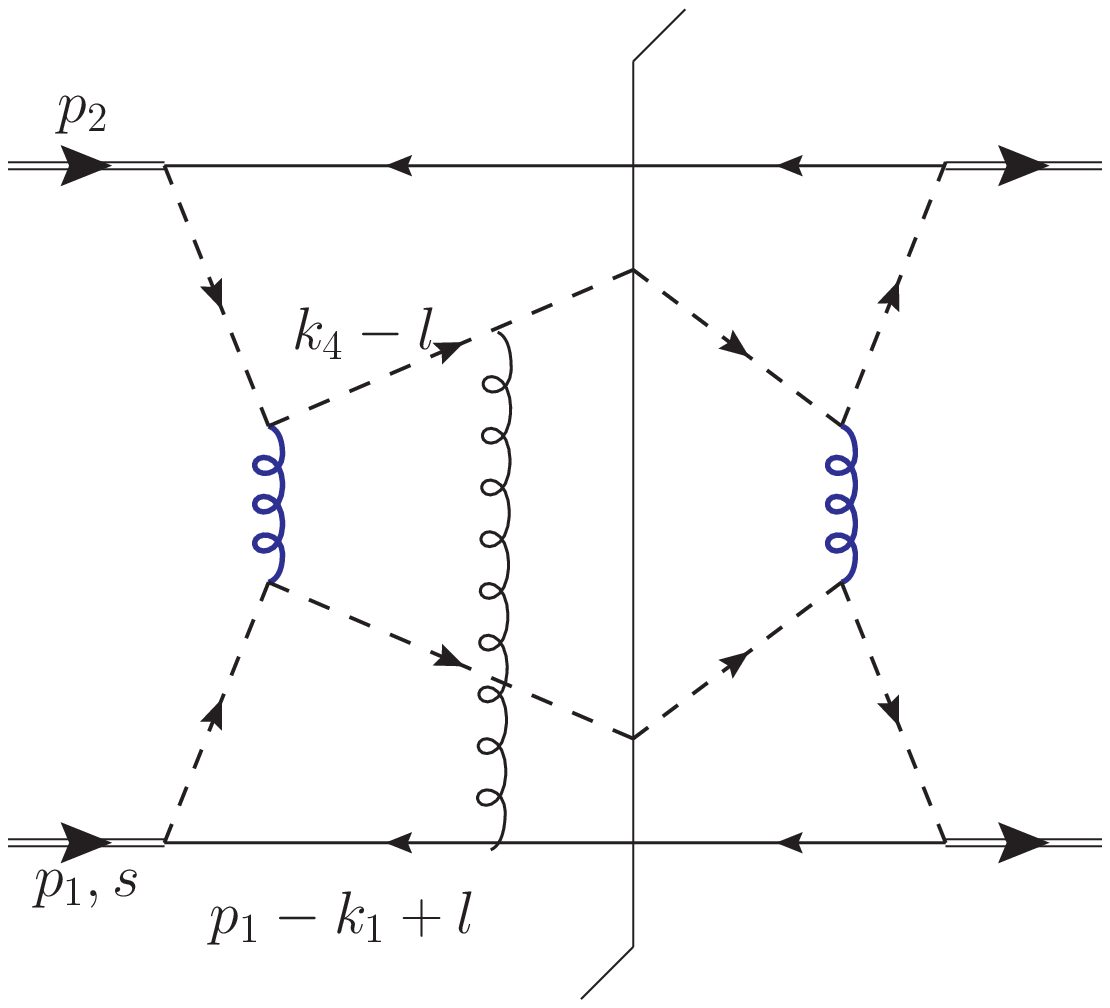}
  &
    \includegraphics[scale=0.45]{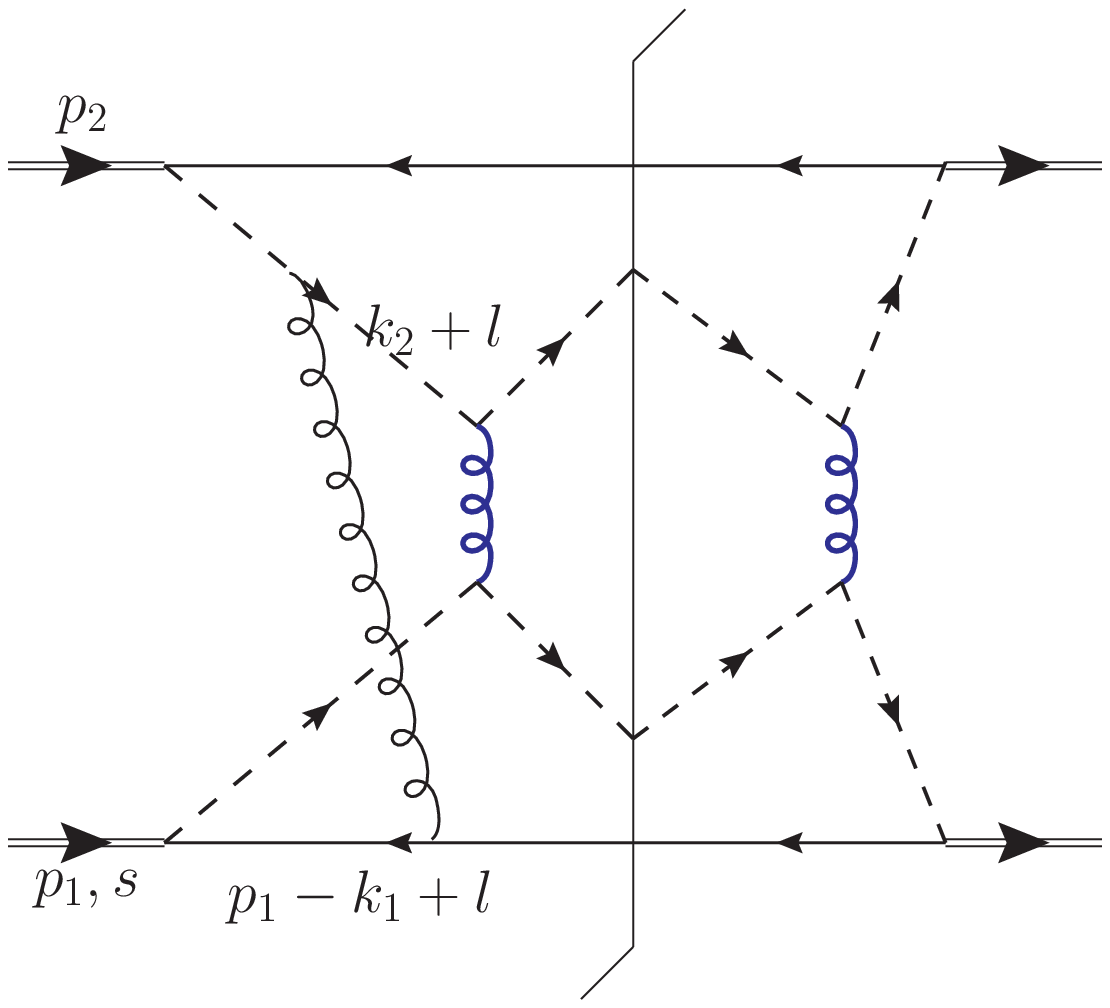}
  &
    \includegraphics[scale=0.45]{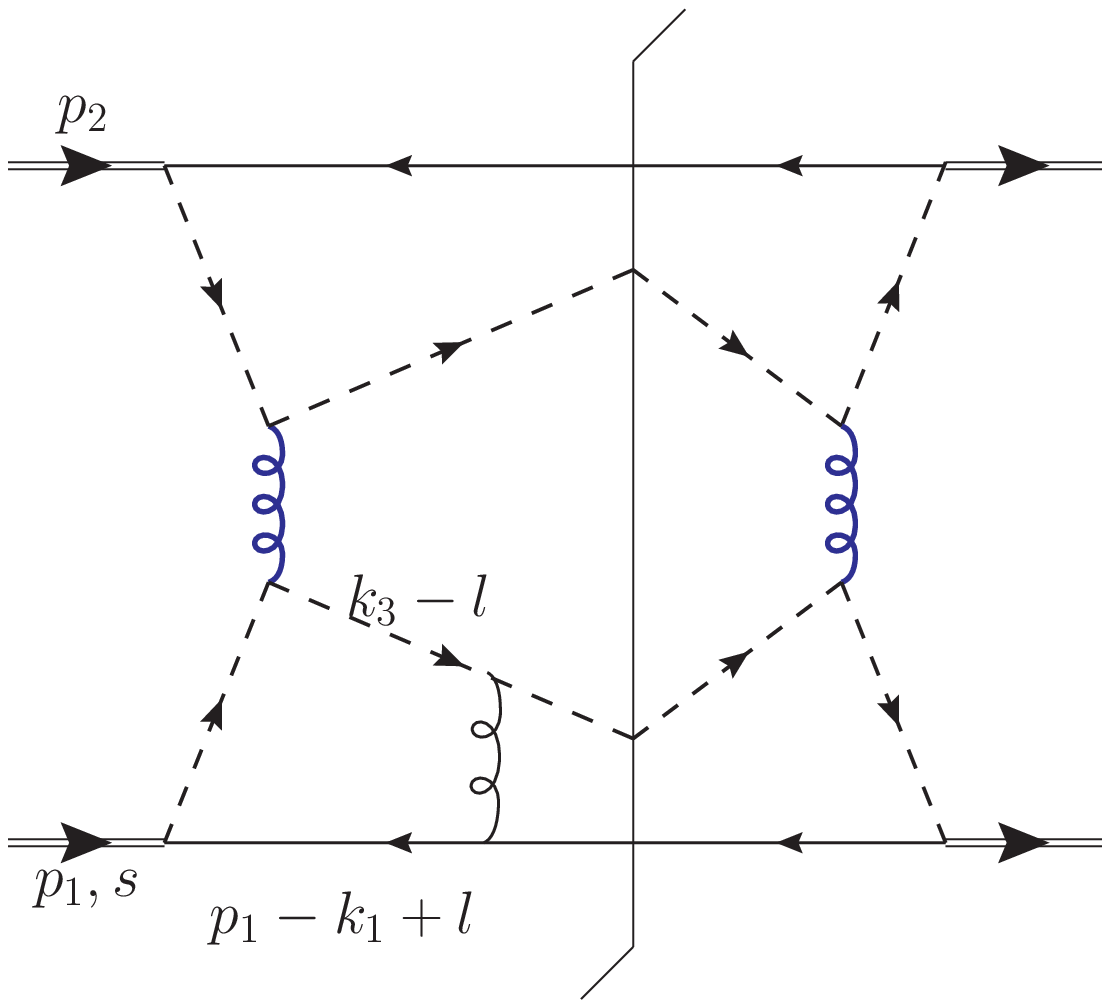}
  \\
    (a) & (b) & (c)
  \end{tabular}
  \caption{One-gluon exchange in model for hadroproduction of hadrons
    of high transverse momentum. The specification of the process is
    as in Fig.\ \ref{fig:HHHH0}.  Only graphs that contribute to the
    SSA are shown.  Hermitian conjugates of these graphs also
    contribute, with an equal value.  }
  \label{fig:HHHH1}
\end{figure*}

The graphs giving the lowest-order SSA are shown in Fig.\
\ref{fig:HHHH1}.  They have an extra gluon exchanged between the
spectator line in the polarized hadron and one of the active partons.
As with the DY process, a sum over cuts of graphs with a
spectator-spectator interaction cancels, while exchanges purely
between active partons give no SSA.  An exchange purely between the
spectator ($p_1-k_1$) and the active parton line ($k_1$) in the
polarized hadron has no relevant intermediate on-shell state and
therefore does not contribute to the SSA.  

So only the three graphs shown in Fig.\ \ref{fig:HHHH1} contribute to
the SSA, together with their conjugate graphs.  Exactly as in our
discussion of SIDIS and DY, the graphs are the same to the leading
power except for eikonal factors from the parton lines connecting the
upper end of the gluon to the hard scattering.  The appropriate
replacements for these lines are
\begin{align}
\label{eq:eik1}
    \frac{ -g_2(2k_4^\mu-l^\mu) }{ (k_4-l)^2 -m_{\phi2}^2+i\epsilon }
  \mapsto {}&
    ig_2\pi\delta_-^\mu \delta(l^+),
\\
\label{eq:eik2}
    \frac{ -g_2(2k_2^\mu+l^\mu) }{ (k_2+l)^2 -m_{\phi2}^2+i\epsilon }
  \mapsto {}&
    ig_2\pi\delta_-^\mu \delta(l^+),
\\
\label{eq:eik3}
    \frac{ -g_1(2k_3^\mu-l^\mu) }{ (k_3-l)^2 -m_{\phi1}^2+i\epsilon }
  \mapsto {}&
    ig_1\pi\delta_-^\mu \delta(l^+),
\end{align}
for a total of $i\pi(g_1+2g_2)\delta(l^+)$.

The $g_1$ term corresponds to a gluon coupling to a future-pointing
Wilson line in the operator definition of the parton density, the same
as for SIDIS.  However, it is impossible for the contribution
proportional to $g_2$ to be represented in terms of a Wilson line
connecting the two parton fields for the distribution of partons of
type $\phi_1$ in the hadron $H_1$.  This is simply because the coupling
for any such Wilson line has to correspond to the color charge of the
parton, i.e., $g_1$, and not $g_2$.  The full Wilson line, or some
generalization, is needed because exchanges of multiple gluons also
contribute.  The quantity we are looking at is
definitely associated with the hadron $H_1$ rather than being in some
kind of exotic soft factor, since the attachment of lower end of the
gluon line to the spectator parton is necessary for the non-zero SSA;
the triviality of this observation is a special feature of our
particular model.  

The contribution to the SSA is therefore non-universal and does not
correspond to a parton density.  That is, factorization is broken.

Of course, the fact that the contribution is obtained from an
eikonalized line indicates that it can be obtained from some kind of
representation in terms of Wilson line operators.  But the matrix
element is for some more complicated and non-universal object 
\cite{bomhof_04,Bacchetta:2005rm,Bomhof:2006dp,Pijlman:2006tq} that
cannot be treated as a parton density.  It is allied to the objects
used by Balitsky \cite{Balitsky:2001gj} to discuss scattering at
high-energy and small angles.  The eikonalization indicates that
substantial simplifications are possible.  But that situation would go
well beyond normal factorization.

\section{Discussion}
\label{sec:discussion}

We should first emphasize that there is a large overlap between the
present paper and the work in Refs.\
\cite{bomhof_04,Bacchetta:2005rm,Bomhof:2006dp,Pijlman:2006tq}.
What is not so clear from the earlier work is whether factorization in
any standard sense continues to hold for in the process
(\ref{eq:HHHH}).  For example, in \cite{Bacchetta:2005rm}, we read 
``We
have assumed factorization to hold in this treatment of
TMD effects although it is, at present, certainly not clear
whether such a factorization holds for hadron-hadron scattering
processes with explicitly TMD correlators.'' 

Our primary result is to show by a counterexample that hard scattering
$k_T$-factorization with universal parton densities fails for the
production of high $p_T$ hadrons in hadron-hadron collisions, when a
pair of measured hadrons is close to back-to-back azimuthally.  The
overall issue is that in a gauge theory arbitrary exchanges of gauge
fields between different collinear groups (``jets'') can occur without
any power suppression.  To obtain factorization it is necessary to
show that the sum over these exchanges can be absorbed into the
definitions of the parton densities and fragmentation functions,
assisted by certain cancellations.  A full proof will be quite
general, applying to a general gauge theory and to many reactions.  So
one particular counterexample is sufficient to show that such a proof
does not exist; we can then choose the counterexample for maximum
clarity and simplicity.

Even for those cases where factorization does hold, the need to make
suitable definitions of the parton densities, etc, so as to absorb the
effects of the gluon exchanges indicates that the parton densities,
etc, can always be regarded as \emph{effective} densities
\cite{Ratcliffe:2007ye}.  The primary practical issue is whether they
are universal, i.e., the same for all reactions.  In a certain sense,
the well-known scale dependence of the densities is a kind of
non-universality: different parton densities are needed when the scale
of the hard scattering is given a large increment.  But there is an
evolution equation for the scale-dependence, and this applies to an
individual parton density.  No details or specification of the hard
scattering is needed to treat the evolution equation, either to derive
it or to apply it.  We should therefore refer in this case to
``modified universality'', not to non-universality.  Similarly the
reversal of the sign of the Sivers function between SIDIS and DY
processes is a case of modified universality.

At the upper end of the exchanged gluon in our counterexample, the
interactions can be treated in the eikonal approximation.  This is
very similar to other discussions of partons passing through the gluon
field of another hadron.  A selection of relevant papers is
\cite{Balitsky:2001gj,Frankfurt:2007rn,Peigne:2002iw,Brodsky:2002ue}.
Much of that work concerns the small $x$ region, diffractive
scattering, etc, whereas our counterexample applies in the fully
conventional region where normal parton-model concepts are generally
considered as fully applicable, i.e., parton fractional momenta are
moderate and the scale of the hard scattering is comparable to the
center-of-mass energy rather than being much less.

Of course, interesting simplifications do occur, so that useful
quantitative estimates can surely be obtained for the non-factorizing
effects.  However the methods are rather different than those for
conventional factorization.  Refs.\
\cite{Balitsky:2001gj,Frankfurt:2007rn,Peigne:2002iw,Brodsky:2002ue}
indicate that the effects of the eikonalized interactions are
substantial, so that the numerical effects of non-factorization should
be significant; in the present paper we did not estimate the numerical
size of the non-factorization.

The gluon exchanges in our counterexample are clearly tied to the
target hadron at their lower end.  But the coupling at the upper end
concerns some parton other than the one coming out of the lower
hadron.  The non-canceling terms are sufficiently tied to the color
flow at the hard interaction that they are not universal in any normal
sense.  This is the clearest indication of non-universality.

\begin{figure}
    \includegraphics[scale=0.45]{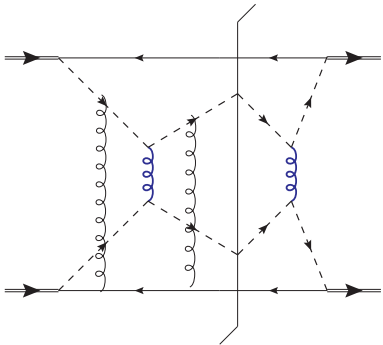}
  \caption{The exchange of two extra gluons, as in this graph, will
    tend to give non-factorization in unpolarized cross sections. }
  \label{fig:HHHH2}
\end{figure}

The reader should not be misled by specific features of our
counterexample into supposing that the failure of factorization is
correspondingly restricted.  These features include: the use of an
SSA, the particular features of the model, and the particular order of
perturbation theory.  The use of the SSA is simply a way of getting
the maximal conceptual sensitivity to problems in constructing a proof
of factorization.  For an unpolarized cross section, we would need an
extra gluon to be exchanged in order for the nonfactorization issues
to arise, from graphs such as those in Fig.\ \ref{fig:HHHH2}.
Evidently, to demonstrate nonfactorization explicitly in this case,
the number of graphs would be larger than in our example, and the
explicit calculations would be much more lengthy.  Standard
power-counting arguments show that the contribution of this and
related graphs is of leading power.  It is very important to determine
whether or not the sum of the potentially non-factorizing
contributions actually does or does not cancel in the unpolarized
cross section.

Similarly, the choice of quantum numbers of the parton fields, of the
abelian gauge group, and of the quantum numbers for the detected
particles was simply to provide maximum transparency and simplicity to
the counterexample.  

The fact that non-factorization can only occur with at least two extra
gluons in a unpolarized cross section might suggest that the
non-factorization is at high order in the strong coupling and
therefore substantially suppressed.  However the region of interest is
at low virtuality for the gluons, so that the appropriate coupling is
for a low momentum scale, where QCD is a strongly coupled theory.

\begin{figure}
    \includegraphics[scale=0.45]{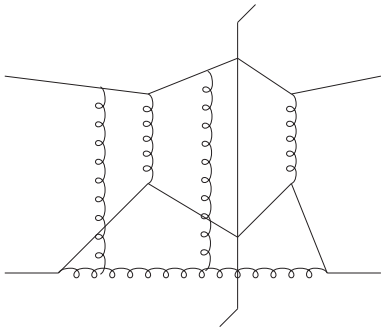}
  \caption{In a conventional perturbative QCD calculation for an
    unpolarized partonic cross section, non-factorization by the
    mechanisms discussed in this paper would first appear in graphs of
    this order. }
  \label{fig:qqg2}
\end{figure}

Even so, the number of extra gluons needed implies that the effects of
non-factorization will only appear in quite high order in conventional
perturbative QCD calculations.  Normally one performs calculations
with on-shell massless quarks and gluons, and extracts collinear
divergences that are grouped with parton densities and fragmentation
functions; any remaining divergences cancel between graphs.
Non-factorization in the hadronic cross section corresponds to
uncanceled divergences in quark-gluon calculations.  The lowest order
in which the mechanisms we have discussed could possible give an
uncanceled divergence in unpolarized partonic cross sections is 
next-to-next-to-next-to-leading-order (NNNLO),
as in Fig.\ \ref{fig:qqg2}.  The region for the uncanceled divergence
is where the lower gluon is collinear to the lower incoming quark, and
two of the exchanged gluons are soft.  This graph is at least one
order beyond all standard perturbative QCD calculations.

Because our calculations directly concern cross sections that use
transverse-momentum-dependent parton densities, a certain amount of
care is needed in interpreting the results.  The natural direction for
the Wilson lines is light-like, as from Eq.\ (\ref{eq:qkl}).  However
light-like Wilson lines give divergences in
transverse-momentum-dependent densities \cite{TMD}.  These are due to
rapidity divergences \cite{C03} in integrals over gluon momentum; they
cancel \cite{TMD} in conventional parton densities only because of an
integral over all transverse momentum in integrated parton densities.
The solution adopted by Collins, Soper and Sterman \cite{TMD} (CSS)
was to define parton densities without Wilson lines but in a
non-light-like axial gauge.  The gauge-fixing vector introduces a
cut-off on gluon rapidity, and then an evolution equation with 
respect to the cut-off was derived.  The non-perturbative functions
involved in this CSS evolution equation have been measured (e.g.,
\cite{Landry:2002ix}) in fits to DY cross sections, and would be an
essential ingredient in testing non-factorization.

However, there are some unsatisfactory features of the use of axial
gauges, which are made particularly evident in polarized cross
sections.  This includes complications concerning gauge links at
infinity \cite{Belitsky:2002sm}, when a Wilson line formalism is used.
A much better definition is to use a non-light-like Wilson line.  This
again obeys an equation of the CSS form.  It is also possible to use a
subtractive formalism \cite{CM,C03} with light-like Wilson lines but
with generalized renormalization factors involving vacuum expectation
values of Wilson lines, which also implement a rapidity cutoff, and
lead to a CSS equation.

To test the predicted non-factorization, we simply need predictions of
high-$p_T$ hadrons in hadron-hadron collisions, made on the basis of
fits to parton densities in DIS and DY and to fragmentation functions
in $e^+e^-$ and SIDIS \cite{Nadolsky:2000ky}.  Probing the SSA would
be particularly interesting, and such measurements are underway at
Relativistic Heavy Ion Collider (RHIC)
\cite{Adams:2003fx,Adler:2005in}.  The same physics is probed in
the transverse shape of jets, and would be worth investigating.

Our counterexample applies in a kinematic region where the normal
intuitive ideas of the parton model appear quite appropriate, even
with a generalization to $k_T$-factorization.  Therefore it forces us
to question under what conditions factorization is actually valid and
to what extent it has actually been demonstrated.  It cannot be
assumed that naive extensions of apparently established results are
applicable beyond the cases to which the actual proofs explicitly
apply.

For hadron-hadron collisions, factorization has been proved
\cite{bodwin_85,collins_85_88} for the Drell-Yan process integrated
over transverse momentum or at large transverse momentum (of order
$Q$).  These proofs apply in the presence of gluon exchanges of the
kind that we discuss in the present paper.  But these papers do not go
beyond this, to the production of hadrons.  Because factorization is
important to all aspects of hadron-collider phenomenology, it is
critical to solve this problem for the hadroproduction of high-$p_T$
hadrons.  Given our counterexample to $k_T$-factorization, a proof of
factorization can only succeed in a situation where conventional
collinear factorization is appropriate.  For dihadron production this
is when the hadron-pair has itself large transverse momentum or when
the pair's out-of-plane transverse momentum is integrated over a wide
range.

In fact, Nayak, Qiu and Sterman \cite{Nayak:2005rt} have recently
given strong arguments that collinear factorization does indeed hold
in such a situations.  The graphs examined are similar to ours.  They
apply Ward identities to prove an eikonalization generalizing our
specific calculations.  Then they observe that a unitarity
cancellation occurs of a kind endemic in factorization proofs
\cite{Collins:1981ta,collins_85_88}.  This concerns graphs that are
related by different placements of the final-state cut.  In our model,
one example is given by Fig.\ \ref{fig:SIDIS1}(a) and (b), and another
is Fig.\ \ref{fig:HHHH1}(a) and its conjugate.  Such cancellations
fail in our examples, because the final-states of the related graphs
have different transverse momentum, and the cross section is not
sufficiently inclusive in transverse momentum.

Mechanisms that cause $k_T$-factorization to fail in back-to-back
hadron production also tend to cause resummation methods to fail.
They will also tend to break factorization or cause large perturbative
corrections when detailed distributions of final-state hadrons are
examined.  Since many such cases are implicit in the analysis of
complicated multi-jet cross sections, and of jet shapes and the like,
encountered in searches for new physics, further understanding is
essential as are quantitative estimates of the effects.  They can have
a particularly important effect in the extrapolation to the LHC of
quantitatively measured distributions at the Tevatron, for example as
embodied in Monte-Carlo event generators.  The methods of
\cite{Balitsky:2001gj,Frankfurt:2007rn,Peigne:2002iw} will be
important.  Probably some of these effects have already been modeled
in some approximation and in at least some Monte-Carlo event
generators, for example by the soft color interaction model
\cite{SCI}.

Troublesome though it may be for phenomenology, breaking of
factorization should be viewed not as some kind of failure, but as an
opportunity.  Examination of the distribution of
high-transverse-momentum hadrons in hadron-hadron collisions will lead
to interesting non-trivial phenomena.

\begin{acknowledgments}
  This work was supported in part by the U.S.\ Department of Energy
  under grant numbers DE-FG02-90ER-40577 and DE-FG02-87ER40371, and
  contract number DE-AC02-06CH11357.  We would like to thank T.
  Rogers, A. Stasto, G. Sterman, M. Strikman, W. Vogelsang, and F.
  Yuan for useful conversations.
  
  The figures in this paper were made using JaxoDraw \cite{jaxodraw}.
\end{acknowledgments}


\end{document}